\documentclass[aps,english]{revtex4}

\usepackage[T1]{fontenc}
\usepackage{textcomp}
\usepackage{amsmath}
\usepackage{graphicx}
\usepackage{SIunits}
\usepackage{babel}

\begin{document}

\title{Strain-gradient mapping of semiconductor quantum dots }

\author{P.-L. de Assis $^{*,1,2,3}$, I. Yeo$^{*,1,2,4}$, A. Gloppe $^{1,5,\dagger}$, H.A. Nguyen$^{1,2}$, D. Tumanov$^{1,2}$, E. Dupont-Ferrier$^{1,5}$, N.S. Malik$^{1,4}$, E. Dupuy$^{1,4}$, J. Claudon$^{1,4}$, J.-M. G\'{e}rard$^{1,4}$, A. Auff\`{e}ves$^{1,2}$, O. Arcizet$^{1,5}$, M. Richard$^{1,2}$, and J.-Ph. Poizat$^{1,2,\ddagger}$}

\affiliation{$^1$ Univ. Grenoble Alpes,  F-38000 Grenoble, France \\
$^2$ CNRS, Inst. NEEL, "Nanophysique et semiconducteurs" group, F-38000 Grenoble France \\
$^3$ Departamento de F\'isica, Instituto de Ci\^encias Exatas, Universidade Federal de Minas Gerais, CP 702, 31270-901, Belo Horizonte, Minas Gerais, Brazil\\
$^4$ CEA, INAC-PHELIQS, "Nanophysique et semiconducteurs" group, F-38000 Grenoble, France, \\
$^5$ CNRS, Inst. NEEL,  F-38000 Grenoble France \\
* These authors contributed equally to this work.\\
$\dagger$, Present address: RCAST, The University of Tokyo, Meguro-ku, Tokyo 153-8904, Japan \\
$\ddagger$ Corresponding author: jean-philippe.poizat@grenoble.cnrs.fr}

\maketitle

\textbf{In the context of fast developing quantum technologies, locating single quantum objects embedded in solid or fluid environment while keeping their properties unchanged is a crucial requirement as well as a challenge to engineer their mutual interaction. Such ``quantum microscopes'' have been demonstrated already for NV-centers embedded in diamond \cite{Rittweger}, and for single atoms within an ultracold gas \cite{Bakr}. In this work, we demonstrate a new method to determine non-destructively the position of randomly distributed semiconductor quantum dots (QDs) deeply embedded in a solid photonic waveguide. By setting the wire in an oscillating motion, we generate large stress gradients across the QDs plane. We then exploit the fact that the QDs emission frequency is highly sensitive to the local material stress \cite{Trotta,Yeo,Montinaro} to infer their positions with an accuracy ranging from $\pm 35\: $nm down to $\pm 1\: $nm.}


Due to their atomic-like properties, semiconductor quantum dots have been widely used for solid-state cavity-quantum electrodynamics~\cite{Gerard,Lodahl}, and display attractive applications in the field of quantum photonics as a key building block to realize bright quantum-light sources~\cite{Claudon,Dousse,Munsch,Ding} and spin-photon interfaces~\cite{Besombes,Awschalom,DeGreve,Gao}. To date, high-quality QDs are mainly obtained via a self-assembly process, which produces nano-islands that are randomly distributed in a plane.
An accurate, non-destructive mapping of the QD position is thus a key capability in the general context of quantum nanophotonics.
The precise location of a QD within a photonic nanostructure indeed
determines the strength of light-matter coupling.
The  relative position of distinct QDs is also crucial to investigate their mutual coupling~\cite{Wieck_2005,Kasprzak,Savona_2005,Mermillod} and collective effects such as superradiance \cite{Temnov,Auffeves}. Although impressive results have already been obtained by all-optical imaging methods~\cite{Betzig_NSOM2,Rittweger,Matsuda}, they are not well suited for imaging quantum dots deeply embedded in photonic nanostructures.


In this work, we propose and demonstrate a non-destructive QD mapping technique whose principle is reminiscent of Magnetic Resonance Imaging (MRI)~\cite{Brown,Mamin,Degen}. By setting the structure in which the QDs are embedded into vibration, we generate across the QD plane an oscillating stress field with a large spatial gradient. The  QD emission energies then display a spatially-dependent oscillation, whose amplitudes and phases are resolved by stroboscopic microphotoluminescence. By using two cross-polarized mechanical modes, we perform a 2D mapping of the QD position in the growth plane. As opposed to optical near-field techniques, this method can be used to determine the position of QDs that are deeply embedded within a solid-state microstructure, with a spatial resolution which is not bound by the laws of electromagnetism.



\begin{figure*}[t]
\includegraphics[width=0.8\textwidth]{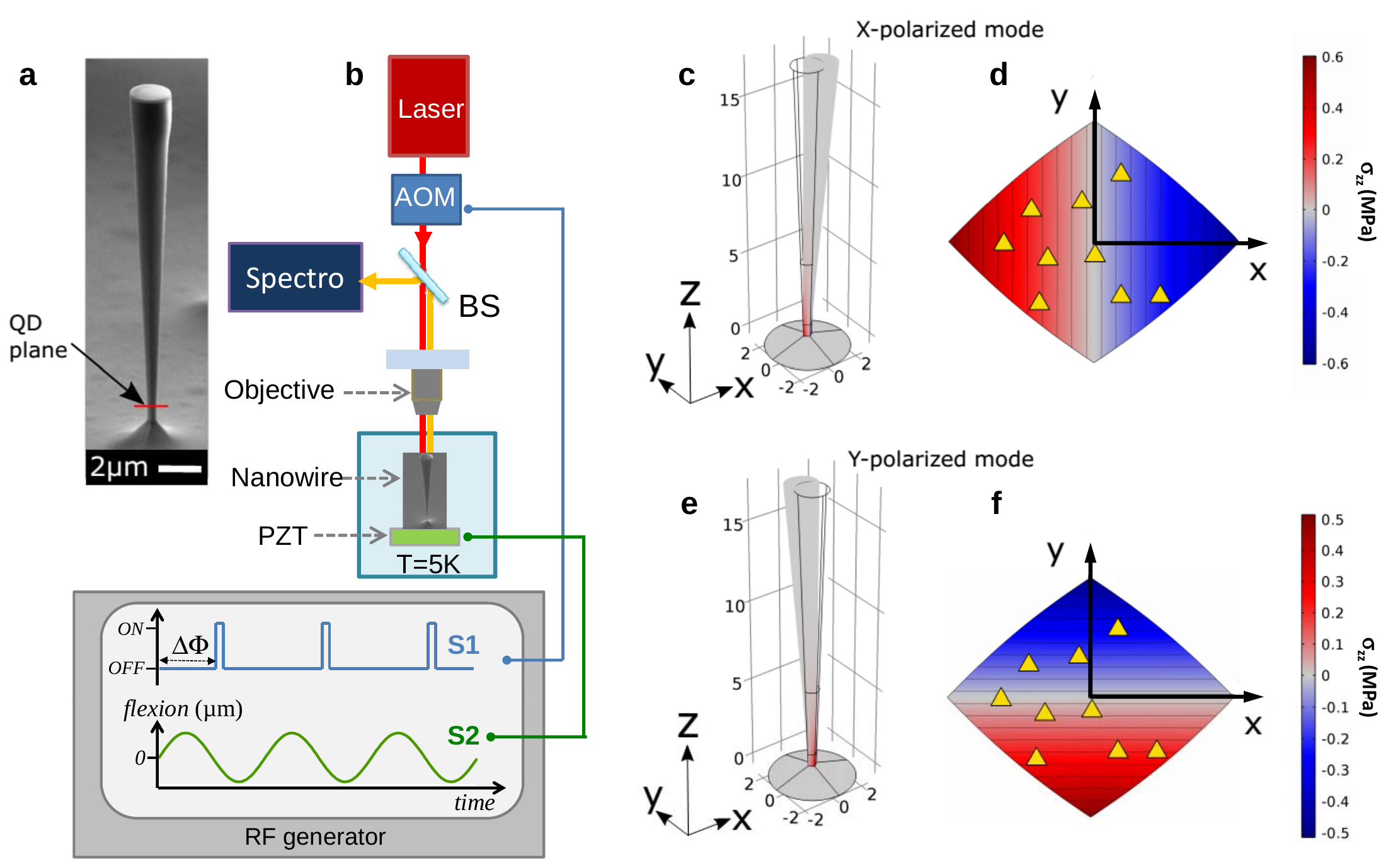}
\caption{\textbf{Principle of strain-gradient mapping and experimental set-up.} \textbf{a}, Scanning electron microscope image of a photonic wire. The line indicates the longitudinal position of the QD plane. \textbf{b}, Experimental set-up. Mechanical modes are excited by a piezoelectrical transducer (PZT) glued on the sample back side. The QD photoluminescence is excited non-resonantly by a laser tuned to $\lambda=\unit{830}{nm}$. To perform stroboscopic measurements of strain-induced QD energy shifts, a radio-frequency (RF) generator simultaneously drives the PZT and an acousto-optic modulator (AOM) which gates the excitation laser. Strain mapping exploits two wire vibration modes: the mode X corresponds to a flexion polarized along $x$ (\textbf{c}) and the mode Y to a flexion along $y$ (\textbf{e}). \textbf{d} and \textbf{f}, Calculated maps in the QD plane of the dominant stress component ($\sigma_{zz}$). The black lines are iso-stress lines. Modes X and Y feature a large stress gradient, oriented along $x$ and $y$ directions, respectively. Depending on their position, QDs (pictured as yellow triangles) will experience different stress modulations. The resulting modulation of their emission energy for modes X and Y is used to infer their spatial location.}
\label{setup}
\end{figure*}


We demonstrate this technique on a tapered GaAs photonic wire antenna which embeds a single layer of InAs quantum dots (Fig.\ref{setup}.a). Similar photonic structures have been used recently to realize bright sources of single-photon~\cite{Munsch}, and hybrid opto-mechanical systems~\cite{Yeo}. The QDs are randomly distributed in a plane perpendicular to the wire $z$ axis, and located $0.8\: \mu$m above the structure base.


Mechanical spectroscopy of the wire vibration modes is conducted in vacuum, at cryogenic temperature ($\text{T}=5\: $K). The wire is set into motion by a piezoelectrical transducer (PZT) attached to the back side of the sample. To perform mechanical spectroscopy, one sweeps the piezo drive frequency, while monitoring the top facet lateral displacement with an auxiliary laser and a split photodiode (SPD) \cite{Sanii,Yeo,Gloppe}. We focus on the fundamental vibration mode, which corresponds to a flexion of the wire with a single node at its basis. Mechanical spectroscopy reveals here two low energy modes X and Y with eigen frequencies $\Omega_\text{Y}/2\pi=\unit{435}{kHz}$ and $\Omega_\text{X}/2\pi=\unit{503}{kHz}$, and a similar quality factor $Q = 1740$. The mode X (Y) corresponds to a vibration along the $x$ ($y$) axis in the laboratory frame.
The observed mode splitting can be traced back to a lozenge shaped cross section of the wire base, with the longest diagonal aligned with the $x=0$ axis. We characterized this shape in detail by focused ion beam (FIB) milling and scanning electronic microscope (SEM) observation as illustrated in Fig.\ref{setup}.e (see SI, section A for details). These nanomechanical measurements are in quantitative agreement with a numerical finite element method (FEM) simulation, which involves a realistic description of the structure geometry.


In the context of QD location measurement, the splitting between modes X and Y is a precious resource: by simply tuning the excitation PZT frequency, one sets the orientation of the stress field gradient in the QD plane along the $x$ or $y$ direction. This feature is illustrated in Fig.~\ref{setup}\textbf{d},\textbf{f}, which show the calculated maps of $\sigma_{zz}$, the dominant component of the stress tensor, for the two vibration modes. Moreover, the conical wire shape leads to very large stress gradient amplitudes: for a $\unit{1}{nm}$ lateral displacement of the nanowire top facet, the stress gradient typically amounts to $\unit{2}{MPa/\mu m}$.
Like in MRI, a large gradient is a crucial asset to obtain a strongly space-dependent QD response. FEM simulations also show that the gradient is constant to better than $10^{-4}$  in the QD layer. Although not mandatory to enable strain-mapping, such uniformity greatly simplifies the data analysis.

\begin{figure}
\resizebox{0.5\textwidth}{!}{\includegraphics{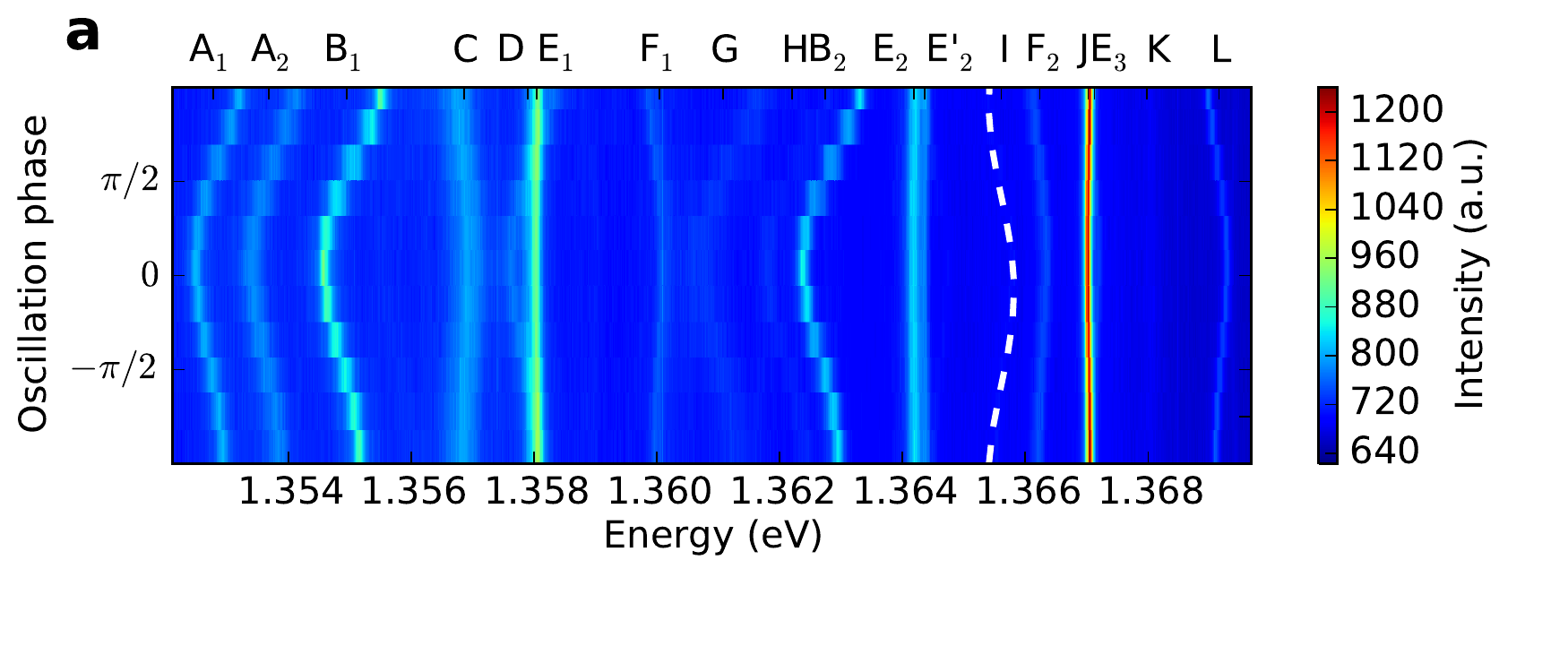}}
\resizebox{0.5\textwidth}{!}{\includegraphics{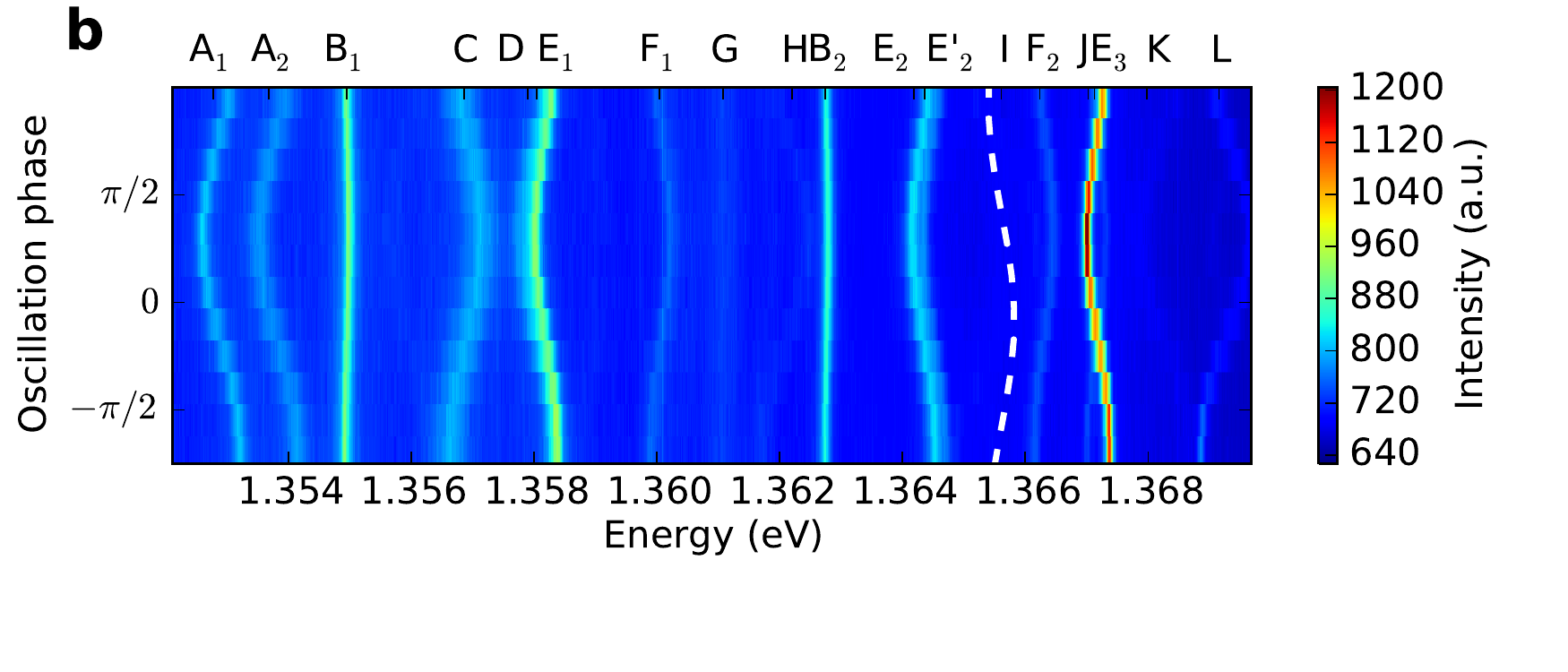}}
\caption{\textbf{Stroboscopic PL for flexural modes X (a) and Y (b)}. The excitation laser intensity is above QD saturation intensity. Lines associated with the same QD are labeled with the same letter and different numeric indices. Emission line I is highlighted due to its very low intensity, by superposing a white dashed sine curve of amplitude, phase and central energy obtained by fitting the values of the central energy of that emission line during an oscillation to a sinusoidal model. Emission lines E$_2$ and E$^\prime_2$ are  likely to be a doublet due to the fine-structure splitting of an exciton, given their very small energy separation, and similar response to wire vibrations. It is possible to see on \textbf{(b)} that QDs E$_3$ and J come into resonance twice during an oscillation.
Note that the phase span is not exactly  $2\pi$ on these two graphs owing to experimental drift during acquisition (cf SI, section F).}
\label{stroboscopic}
\end{figure}

In order to detect the spatially dependent response of the QDs in this oscillating stress field, we realize a stroboscopic optical excitation and detection of the QDs photoluminescence (PL). As shown in Fig.\ref{setup}.b, the radio-frequency generator delivers two RF signals locked in frequency and phase. The first one is harmonic and drives the wire mechanical motion via the PZT; its frequency is chosen to match the X or Y mechanical resonance. The second one is used to gate the laser which excites the QDs PL, and features an ON time $10$ times shorter than the mechanical period. The relative phase $\Delta\phi$ between both signals can be continuously tuned between $0$ and $2\pi$ and the resulting stroboscopic QD PL is sent into a high resolution optical spectrometer. By scanning $\Delta\phi$, one obtains series of PL spectra showing how each QD emission frequency is modulated according to the time-dependent stress that they experience. This measurement is carried out for both mechanical polarization directions M = X and Y; the results are shown in Fig.~\ref{stroboscopic}.

Upon excitation of the mode M, the  emission energy $\hbar \omega_{i}$ of a given spectral line $i$ behaves like
\begin{equation}
\hbar \omega_{i,\text{M}}(\Delta\phi)= \hbar \omega_{i,0}+ \hbar \Delta\omega_{i,\text{M}}\cos(\Delta\phi+\phi_{i,\text{M}}),
\end{equation}
where $\hbar \omega_{i,0}$ is the unperturbed emission energy. The measurement of the modulation amplitudes $(\Delta \omega_{i,\text{X}} ; \Delta \omega_{i,\text{Y}})$ and of the relative phases $(\phi_{i,\text{X}} ; \phi_{i,\text{Y}})$ yields an absolute spatial localization of the QD associated with the spectral line $i$.

As evidenced in Fig.~\ref{stroboscopic}, the  phase $\phi_{i,\text{M}}$  can take two values: $0$ or $\pi$. This phase provides a `which side' information on the location of the emitter with respect to the zero-stress line of mode M. Thus, the couple $(\phi_{i,\text{X}} ; \phi_{i,\text{Y}})$ allows us to place the QD unambiguously in one of the four quadrants defined by the zero-stress lines of modes X and Y (i.e., the $y=0$ and $x=0$ diagonals of the wire section).

In addition, the absolute distances $(|x_i|;|y_i|)$ with respect to the zero-stress lines can be inferred from the frequency modulation amplitudes $(\Delta\omega_{i,\text{X}};\Delta\omega_{i,\text{Y}})$ through the relations
%
\begin{align}
|x_i| &= \hbar \Delta\omega_{i,\text{X}}/(g_i s_\text{X})\\
|y_i| &= \hbar \Delta \omega_{i,\text{Y}}/(g_i s_\text{Y})
\end{align}
We have introduced $g_i \simeq \frac{\text{d} \hbar \omega_i}{\text{d} \sigma_{zz} }$, the tuning slope of transition $i$. Note that the main contribution (more than $97\%$) to the QD energy shift is caused by the $\sigma_{zz}$ component of the strain tensor, since the  $\sigma_{xx}$ and $\sigma_{yy}$ components are already one order of magnitude smaller, and that moreover the QD energy shifts caused by these two latter components are $3.7$ times smaller than for the $\sigma_{zz}$ component (see section E of the SI).
The quantities $s_\text{X} = \big(\frac{\text{d} \sigma_{zz}}{\text{d} x}\big)_\text{X}$ and $s_\text{Y} = \big(\frac{\text{d}\sigma_{zz}}{\text{d} y} \big)_\text{Y}$ are the (constant) in-plane strain gradients of modes X and Y corresponding to the top facet displacements $d_\text{X}$ and $d_\text{Y}$ respectively.  They are given by $s_\text{X} = \mu_\text{X} d_\text{X}$ and $s_\text{Y} = \mu_\text{Y} d_\text{Y}$, where $\mu_\text{X}$ and $\mu_\text{Y}$ are the stress gradient  per top facet displacement computed by FEM. Note that, owing to the wire shape anisotropy, we have $\mu_\text{X}/ \mu_\text{Y}=0.98$. For the data shown in this work, we have $d_\text{X} / d_\text{Y} = 0.87$.


To use this relation, the knowledge of each $g_i$ is in principle required. In the present configuration (local stress applied along the QD growth axis) it has been shown experimentally by some of us \cite{Stepanov} that the relative dispersion $\sqrt{\delta g_i^2}/g_0$ is smaller than  $13 \%$. We will thus assume the same $g_i$ for all emitters, equal to the mean value $g_0$, and include the $g_i$'s variations in the location uncertainty analysis.

Interestingly, if the structure embeds a sufficient number of distinct QDs (say N > 10), mapping does not even require an {\it a priori} knowledge of $g_0$. The mapping procedure then consists in two steps. One first determines the QD relative positioning using stroboscopic data (Fig.\ref{stroboscopic}).
For the relative scaling of the $x$ and $y$ axes, the  top facet vibration amplitudes ratio $d_\text{X}/d_\text{Y}$ of both $X$ and $Y$ polarizations is carefully measured. This allows us to express the frequency shift ratio $\Delta ^0 \omega_{i,\text{X}}/\Delta ^0 \omega_{i,\text{Y}} $ corresponding to the same stress gradient on both $X$ and $Y$ axis as a function of the measured frequency shifts :
$\Delta ^0 \omega_{i,\text{X}}/\Delta ^0 \omega_{i,\text{Y}} = (\mu_\text{Y}d_\text{Y}/\mu_\text{X}d_\text{X})\Delta\omega_{i,\text{X}}/\Delta\omega_{i,\text{Y}}$.
In a second step, one determines the overall scaling factor between the relative map and the actual section geometry. We have checked (see section C of the SI) that QDs are optically active over the whole GaAs wire section. Knowing the shape of the waveguide section, we use as a reference the QD which is the closest to the wire sidewall.  As described in section D of the SI, the normalized distance of this extremal QD to the sidewall can be determined by a statistical argument and is given by $1/(2N+1)$. The standard deviation associated to this estimation scales as $1/(2N)$.



\begin{figure}
\resizebox{0.52\textwidth}{!}{\includegraphics{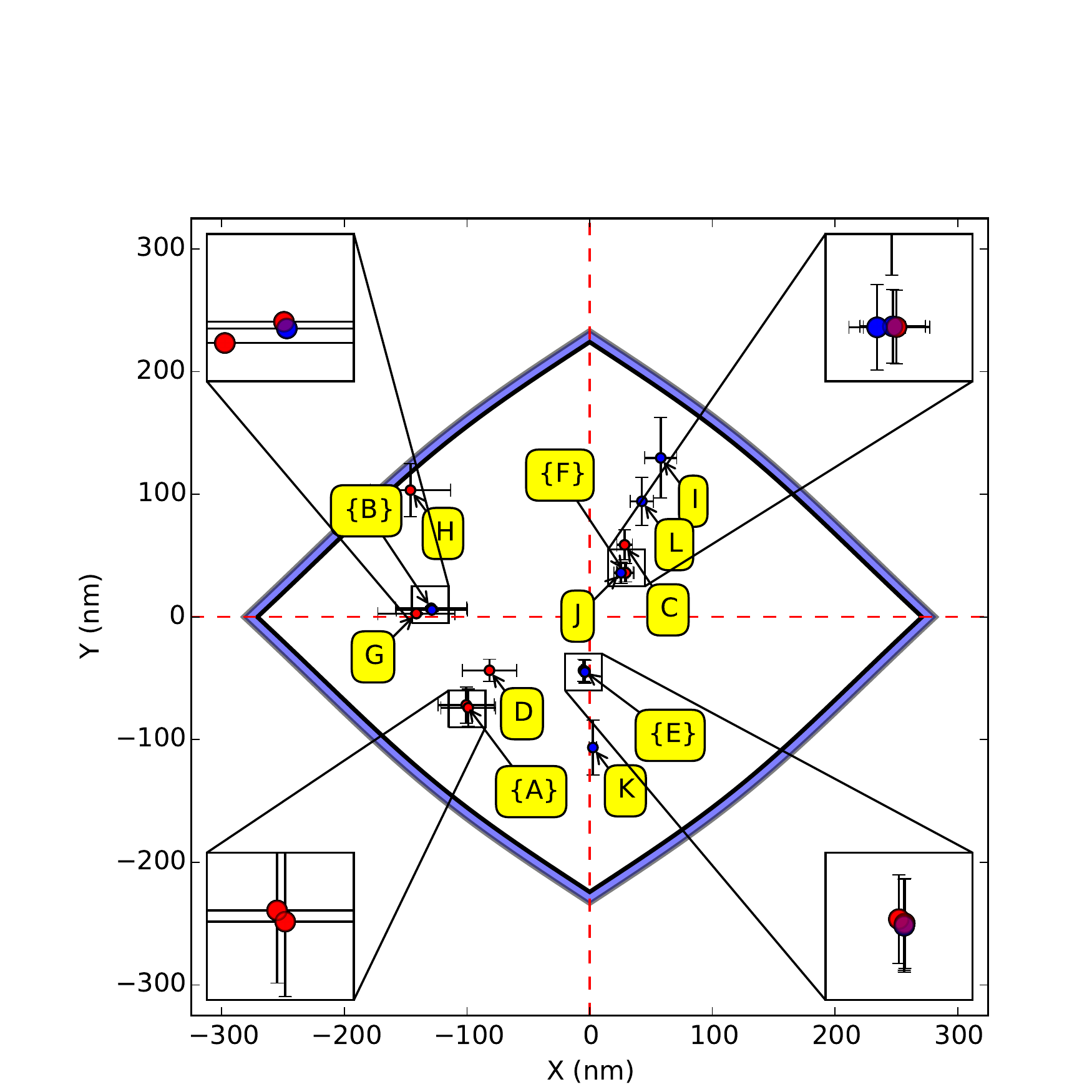}}
\caption{\textbf{QD spatial map} indicating the position inside the nanowire associated with each emission line observed in Fig. \ref{stroboscopic}.  Emission lines with sub-linear and linear (super-linear) power dependence are indicated by red (blue) disks. Error bars are calculated by taking into account the uncertainties of each parameter, for both $x$ and $y$ position. Insets  highlight lines that are found to originate from regions with high overlap. Each inset box is a square with $\unit{30}{nm}$ of side. The QD labeled H is the one closest to the edge and is  used to set the position scale for all QDs (see SI). The outer blue area is a dead layer where QD close to the surface are not optically active (see SI).}
\label{map}
\end{figure}

This procedure allows us to construct the map presented in Fig.\ref{map}, where the labels are consistent with those used for the spectral measurements shown in Fig.\ref{stroboscopic}. The insets show that several emission peaks can be found at the same location. Indeed, a single QD can be responsible for one, two or three emission peaks corresponding to the neutral exciton, the charged exciton and the biexcitonic transitions \cite{Zhiming}. These peaks can be discriminated from each other using their excitation power dependence as described in section B of the SI. When a peak is identified as related to a biexcitonic transition, it always  coexists with a neutral or charged exciton peak within the same dot (i.e. at the same position).
Note that in this analysis, we have implicitly assumed an identical tuning slope for all three aforementioned transitions. This is indeed correct since for such weak strain fields, the strain induced conduction and valence band shifts dominate over the correction of the many-body Coulomb interaction \cite{Kuklewicz,Wu}. With this analysis we could determine that $N = 12$ distinct QDs are present in the QDs plane and within our detection spectral window.

In this mapping procedure, the location of the QDs is found within a certain accuracy, which is fixed by the numerical analysis of the data, as well as by the various assumptions mentioned previously. They can be summarized into several contributions: the measurement accuracy of $\Delta\omega_{i,\text{M}}$ from the dataset of Fig.\ref{stroboscopic}, the statistical determination of the overall scaling factor, the measurement error on the motion amplitude ratio $d_\text{X}/d_\text{Y}$ that impacts the relative scaling on the $x$ and $y$ axes, the non perfectly constant stress gradient $s_\text{M}(x,y)$, and finally, the non-exactly equal response $g_i$'s of each QDs to an equal stress. A detailed analysis of these contributions on the location accuracy of the QDs is given in section F of the SI.
Note that while the total relative uncertainties  vary little between QDs, the absolute uncertainties will be smaller for QDs closer to the neutral lines  (see Fig.\ref{map}).
With this non-destructive method, we are able to determine the QDs positions  inside the nanowire  with an uncertainty as small as $\pm1\;$nm, when it is close to a neutral line, and of at most $\pm35\;$nm, close to the edges. This is to be compared to the diameter on the order of $10$ nm of a QD. This highest uncertainty value corresponds only to $13\%$ of the QD emission wavelength.  Strikingly, the smallest uncertainty corresponds to as little as $0.3\%$ of the QDs emission wavelength.

It turns out that the largest source of uncertainty comes from the dispersion in the response $g_i$ of each QD to a given stress. However, it would be possible, with a different set-up, to measure each $g_i$ independently by exciting the longitudinal breathing mode (at $\sim 30$ MHz) which features a uniform strain within the QD plane. The measured energy shift of each QD$_i$ would then provide us with an \emph{in situ} measurement of $g_i$. Within realistic assumptions, this would reduce the errors bars down to $\pm0.5$ nm close to the center and $\pm20$ nm close to the edge.

Beyond QD devices, this very accurate non-destructive imaging method generalizes to any solid state quantum emitter sensitive to strain such as color centers  like NV centers \cite{Teissier} or rare earth ions \cite{Thorpe}.
The knowledge of quantum emitters positions with respect to each other, as well as with respect to a fixed point in space, opens up new possibilities to characterize and exploit many body quantum optics with such objects, as well as their interaction with nearby surfaces or nanostructures, engineered or unwanted.

\section*{Methods}

\subsection*{Structure fabrication and geometry}
The GaAs photonic wire used in this experiment was fabricated from a planar sample grown by molecular beam epitaxy over a GaAs [001] wafer. The photonic wires are defined by a top-down approach, which employs in particular e-beam lithography and a carefully optimized dry-etching step, conducted in a reactive ion etching chamber \cite{Munsch}. From the top facet to the base waist, the length of the GaAs conical wire is \unit{18}{\micro m}. The top facet is circular and features a diameter of $\unit{1.88}{\micro m}$. Slightly different etching rates along different crystallographic directions result in a base which features a lozenge shape. Its geometry was determined by FIB milling and SEM observation (see section A of the SI). Its major diagonal, oriented along the $x$ direction, features a length of $530\: $nm. The minor diameter is oriented along $y$, with a length of $420\: $nm. This shape anisotropy splits the first flexural mode into two mechanical modes X and Y of orthogonal polarization directions. The wire embeds a single layer of self-assembled InAs QDs, which is located $\unit{0.8}{\micro m}$ above the bottom end of the conical wire; the structure typically embeds a few tens of quantum dots. The wire sidewalls are covered by a $30\: $nm thick Si$_3$N$_4$ passivation layer~\cite{Yeo_2011}, and the top facet is covered by a $170\: $nm thick Si$_3$N$_4$ layer.

\subsection*{Mechanical spectroscopy}
In order to excite and characterize the nanowire vibration, a piezoelectrical transducer (PZT) is placed at the back side of the sample-holder. The PZT is driven by a  sinusoidal voltage, the frequency of which can be scanned across the mechanical resonance. Mechanical spectroscopy of the wire motion is realized by illuminating the wire top facet with a laser, and by measuring the reflected beam deviation with the difference signal from a split photodiode (SPD)  \cite{Yeo,Sanii,Gloppe}.  The polarization of these mechanical modes is experimentally determined by rotating the nanowire top facet image with a Dove prism, prior to detection with the SPD.

\section*{Acknowledgements}
The authors wish to thank E. Gautier for the FIB cut and images. Sample fabrication was carried out in the \textit{Upstream Nanofabrication Facility} (PTA) and CEA LETI MINATEC/DOPT clean rooms. P.-L. de Assis was financially supported by the ANR-WIFO project and CAPES Young Talents Fellowship Grant number 88887.059630/2014-00, and D. Tumanov by the Rh\^{o}ne-Alpes Region.

\newpage

\section*{Supplementary information}

\subsection{Geometry of the mechanical oscillator}\label{Cross-section}

The fact that the first flexural mode of the photonic wire is split into two orthogonal directions with significantly different frequencies indicates that there is an intrinsic asymmetry to its shape. This is confirmed by SEM images of a wire cut by a focused ion beam (FIB), indicating a cross-section at the anchorage that can be well approximated by a lozenge with long semi-axis $X_0$ and short semi-axis $Y_0$, so that $Y_0/X_0 =\epsilon$ (see Supplementary Fig.\ref{fib}).

\begin{figure}[b]
\resizebox{0.45\textwidth}{!}{\includegraphics{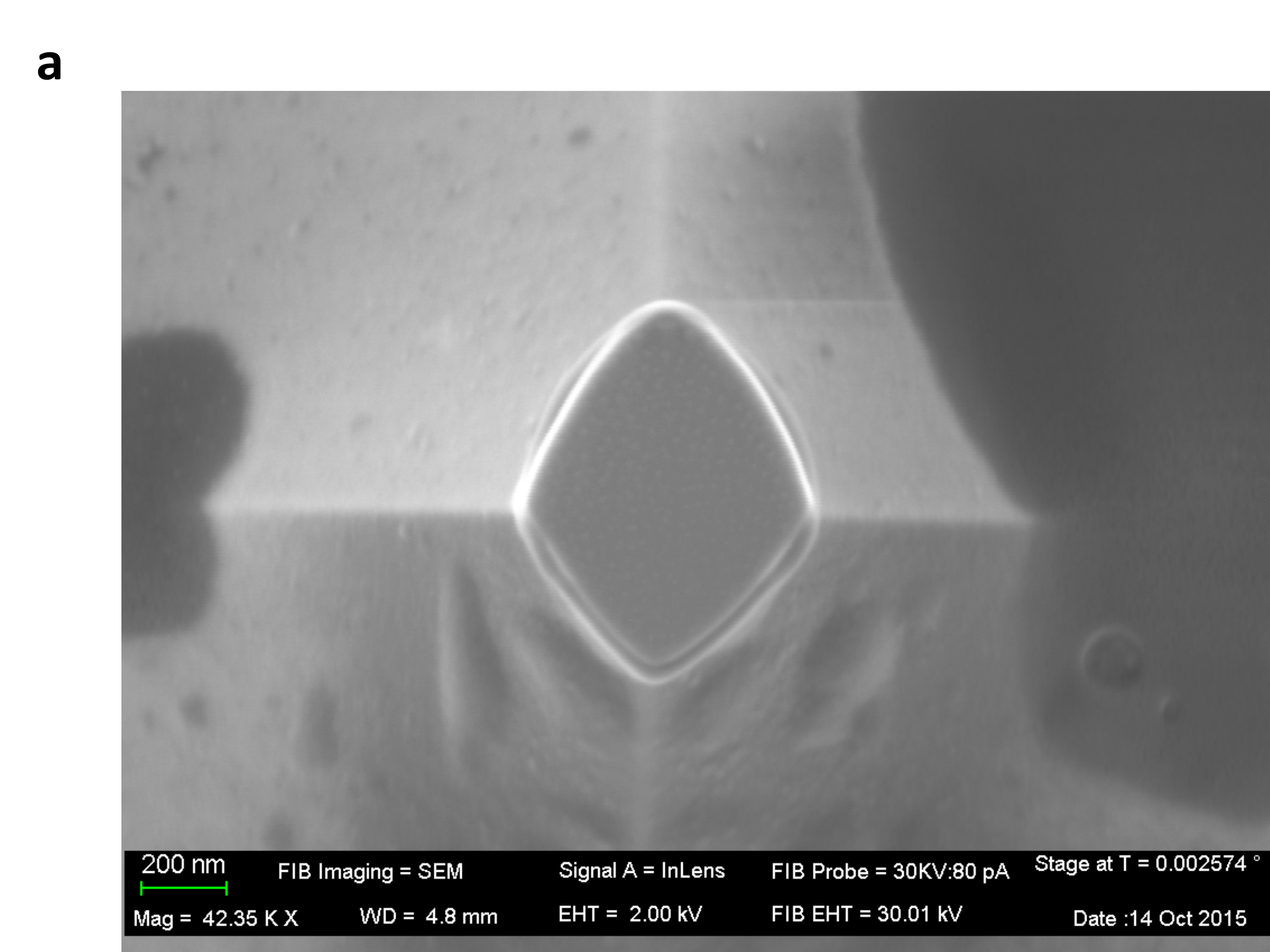}}
\resizebox{0.45\textwidth}{!}{\includegraphics{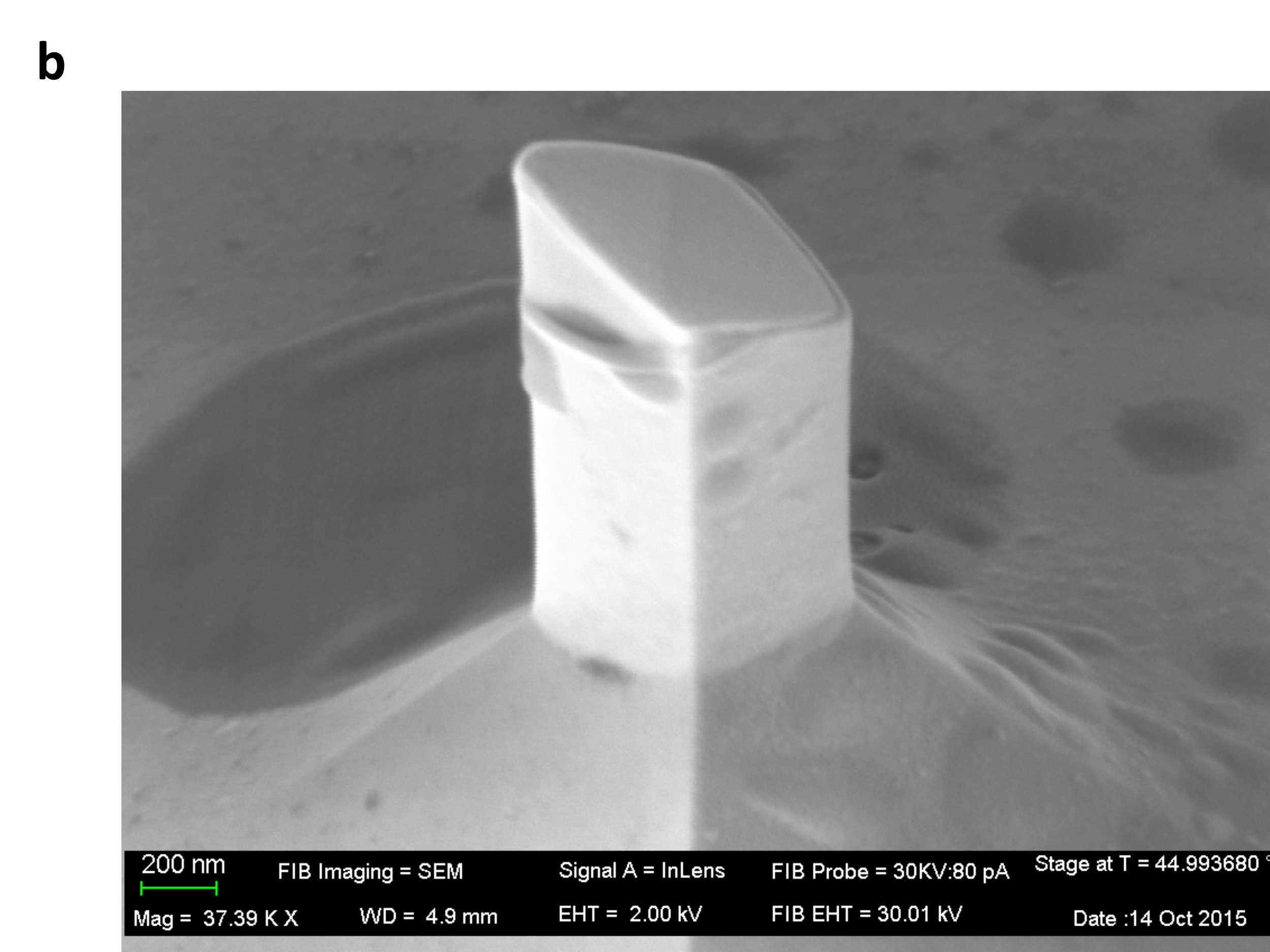}}
 \caption{Scanning electron microscopy images of a typical photonic wire, cut by FIB. In \textbf{a} it is possible to see the lozenge cross-section, with a major and a minor axis, responsible for the broken degeneracy of flexural modes. The seeming asymmetry between the top and bottom part of the lozenge is due to the cut being slanted, as evidenced by the side view in \textbf{b}.}
 \label{fib}
\end{figure}

The transition from a lozenge at the bottom to a circle of radius $R$  at the top of the nanowire ($R=1.88\mu$m) means that at the level of the QDs the cross-section has a shape that can be determined by connecting every point on the bottom ($z=0$) to its corresponding point on the top of the wire (at $z=H=18 \mu$m) by a straight line. This is done by parametrizing both the lozenge and the circle in polar coordinates, as functions of $\theta$, so that for each line both start and end points have the same $\theta$. Thus, for any given position in the $z$ axis it is possible to determine the shape of the cross-section, which we call a bulged lozenge.

At $z=0$, the coordinates of the wire section are given by
\begin{eqnarray}
X_L(\theta) &=& \frac{Y_0}{S_Y(\theta)\tan(\theta)+S_X(\theta)\epsilon}  \\
Y_L(\theta) &=& \frac{Y_0}{S_Y(\theta)+S_X(\theta)\epsilon/\tan(\theta)},
\end{eqnarray}
with
\begin{align}
S_X(\theta)=\begin{cases} +1\ \mbox{if}\ -\pi/2\leq\theta\leq\pi/2   \\
-1\ \mbox{if}\ \pi/2<\theta<3\pi/2
\end{cases}
\end{align}
and
\begin{align}
S_Y(\theta)=\begin{cases}  +1\ \mbox{if}\ 0\leq\theta\leq\pi   \\
-1\ \mbox{if}\ \pi<\theta<2\pi
\end{cases}
\end{align}
At an elevation $z$, the wire section is given by
\begin{eqnarray}
x(\theta,z) &=& \frac{R\cos(\theta)-X_L(\theta)}{H}z+X_L(\theta) \\
y(\theta,z) &=& \frac{R\sin(\theta)-Y_L(\theta)}{H}z+Y_L(\theta) \\
\rho(\theta,z) &=& \sqrt{x^2(\theta,z)+y^2(\theta,z)}.
\end{eqnarray}

Using data from the FIB cut images and finite element simulations to compare theoretical and measured mechanical frequencies, we estimate $X_0=\unit{252}{nm}$ and $Y_0=\unit{201}{nm}$, so that $\epsilon\approx0.8$. The QD plane is at $z=\unit{0.8}{\micro\metre}$, as determined during the epitaxial growth of the sample used to fabricate the photonic wires.

\subsection{Identification of emission lines}
\label{lineID}

As seen in Fig.3 of the main text, many emission lines correspond to spatial location overlapping in the photonic wire. This is due to the fact that excitonic, biexcitonic and trionic lines of the same QD appear
with the laser power used to perform the stroboscopic experiment.

In order to determine which emission lines correspond to excitons and which correspond to biexcitons, we recorded a series of photoluminescence spectra using a wide range of excitation-laser powers. Each line mapped in Fig.3 of the main text was then analysed with respect to the dependence of their integrated intensity with excitation power, $I \propto P^\alpha$. Excitons exhibit $\alpha \leq 1$, with sub-linearity being attributed to charge trapping and other effects that hamper radiative recombination. It is important to note that charged excitons may also exhibit a sub-linear power dependence. Biexcitons, on the other hand, are associated with $1\leq\alpha\leq2$; the inequality being due to the same kind of effects that cause a sub-linear power-dependence for excitonic emission intensity.

In Supplementary Fig.\ref{power}  we show how the emission intensity depends on excitation power for the families of emission lines highlighted in the insets of Fig.3 of the main text. These families of lines are also indicated in Fig.2 of the main text.

\begin{figure}
\resizebox{0.53\textwidth}{!}{\includegraphics{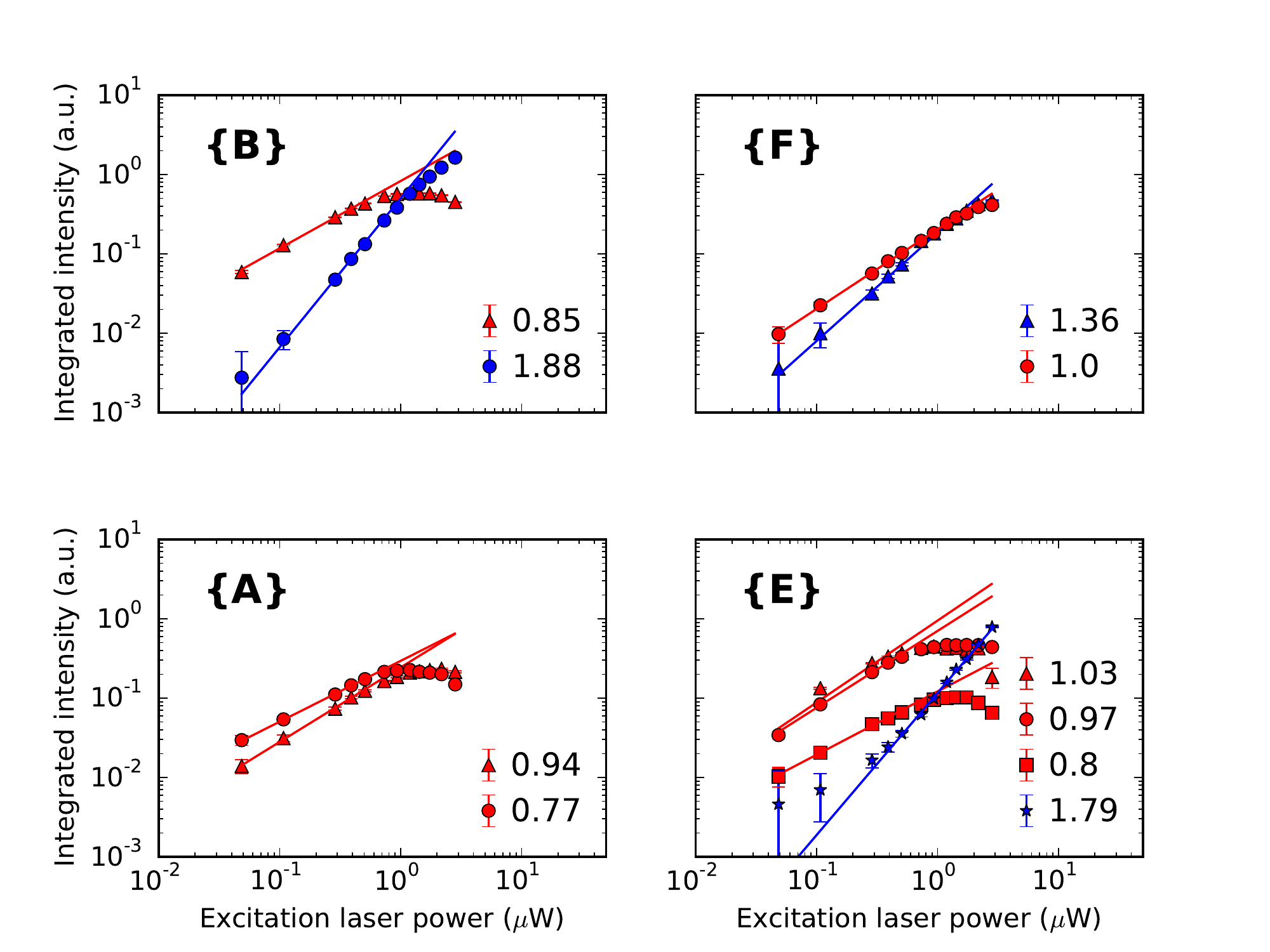}}
 \caption{Excitation power dependence behaviour of emission-line families on the corresponding insets of Fig.3 of the main text, also indicated on Fig.2 of the main text.
 Emission lines with sub-linear and linear behaviour are indicated by red markers, those with  super-linear behaviour are indicated by blue markers.
 For the sake of completeness the values of the exponent $\alpha$ for the different lines are given on the graphs.}
 \label{power}
\end{figure}

\subsection{Determination of the ``dead layer'' thickness}
\label{Dead_layer}

 The wire etching process introduces defect close to its surface.
 Light emission of QDs located close to the outer surface might be suppressed owing to non-radiative recombinations caused by the  surface proximity. Geometrically, such a  ``dead layer'' corresponds to all points in the QD plane that are at a distance $d\leq l$ from at least one point on the wire boundary. The measure of this dead layer thickness $l$ is important to know the effective area where active QDs are located (see  section \ref{scaling}).

This effective boundary will have the same general shape as the outer boundary, with a scaling factor given by the ratio between $x_{DL}$ and $X_0$, where $x_{DL}$ is the point on the $x$ axis that is at a distance $l$ from two points on the outer boundary (see Supplementary Fig.\ref{geometry}).  Given the non-circular nature of the cross section, $x_{DL}<X_0-l$ and is given by
\begin{align}
x_{DL}=\rho(\theta_c)\cos(\theta_c)-l\cos(\alpha(\theta_c)),
\label{Eq_x_DL}
\end{align} where $\pm\theta_c$ is the angular position of the two points tangent to the circle of radius $l$ centered on $x_{DL}$ in Supplementary Fig.\ref{geometry}. In order to properly determine $x_{DL}$, we have to solve
\begin{align}
\rho(\theta_c)\sin(\theta_c)=l\sin(\alpha(\theta_c)),
\label{theta_crit}
\end{align}where $\alpha(\theta)$ is the angle that the vector normal to the boundary at a point $\rho(\theta)$ makes with the $x$ axis:
\begin{align}
\alpha(\theta)=\arctan\left(\frac{\rho(\theta)\tan(\theta)-\frac{\mathrm{d}\rho(\theta)}{\mathrm{d}\theta}}{\rho(\theta)+\tan(\theta)\frac{\mathrm{d}\rho(\theta)}{\mathrm{d}\theta}}\right).
\label{normal}
\end{align}

\begin{figure}
\resizebox{0.5\textwidth}{!}{\includegraphics{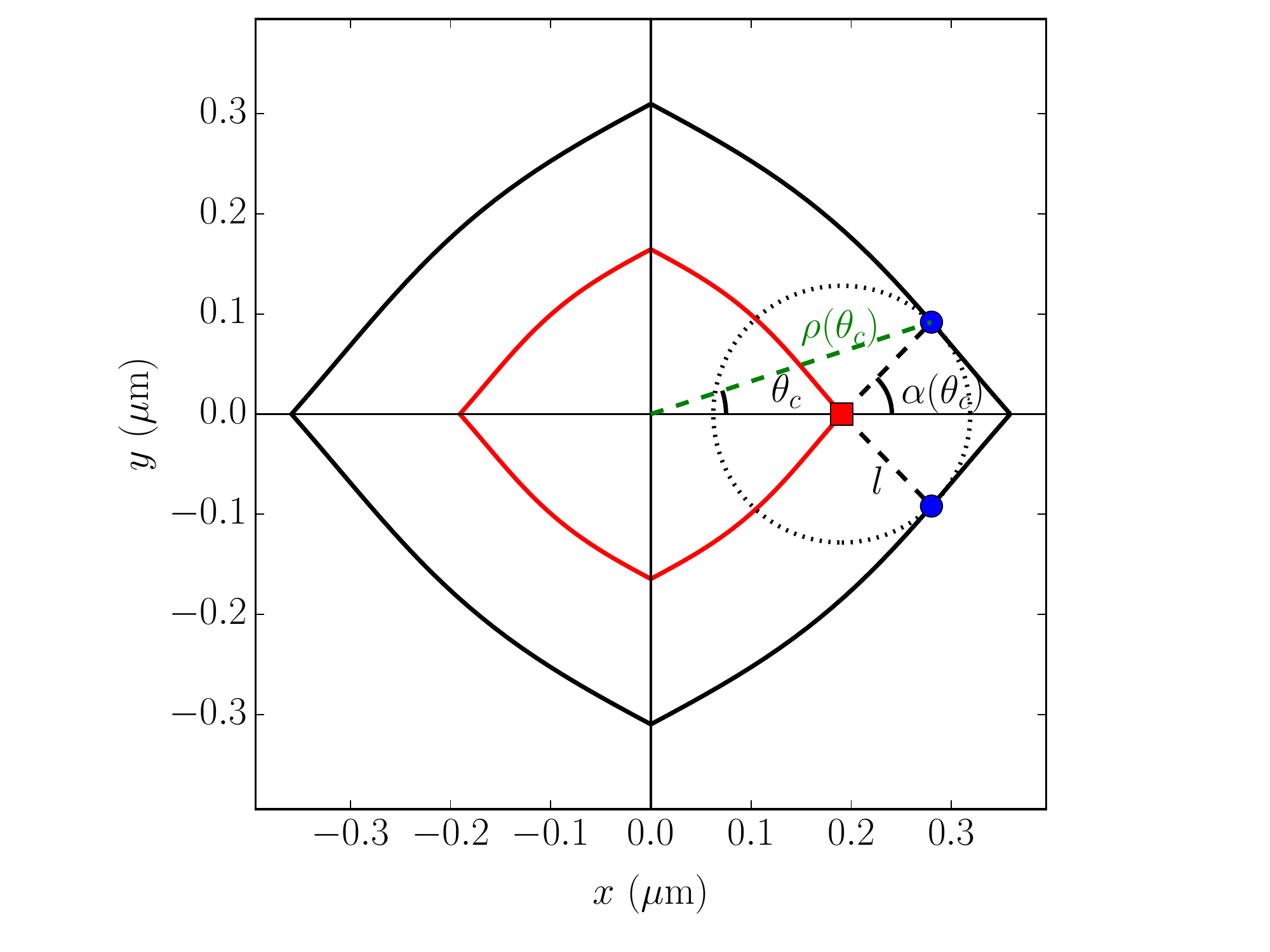}}
 \caption{Geometric construction of the inner boundary of a ``dead layer'' of thickness $l$ on the photonic wire cross section at $z=\unit{2}{\micro\metre}$, drawn as a red continuous line. We wish to determine the position $x_{DL}$ of the extremal point of the inner boundary on the $x$ axis, marked on the figure with a red square. Given the symmetry of the cross section and the definition of the ``dead layer'', a circle of radius $l$ centered on $(x_{DL},0)$, shown as a dotted line, is tangent to the outer boundary on two points, marked by blue dots. $\rho(\theta_c)$ (shown in green) and $\theta_c$ are the radial and angular position of the tangent points, and $\alpha(\theta_c)$ is the angle between the normal to the outer boundary at point $\rho(\theta_c)$ and the $x$ axis, calculated using Eq. \ref{normal}. The position of $x_{DL}$ is determined by solving Eqs. \ref{theta_crit} and \ref{Eq_x_DL}. The value $z=\unit{2}{\micro\metre}$ and a large value of $l$ were chosen instead of the actual values in order to make the illustration clearer.}
 \label{geometry}
\end{figure}

The thickness of this ``dead layer'' is inferred by counting the number of emission lines within a given spectral window for wires with different top radii $R$ (see Supplementary Fig.\ref{Stat}). We have taken photoluminescence spectra of wires with different diameters, using the same spectral window and pumping non-resonantly above the saturation power of the faintest QDs, to assure that none would be ignored due to insufficient excitation.

\begin{figure}
\resizebox{0.4\textwidth}{!}{\includegraphics{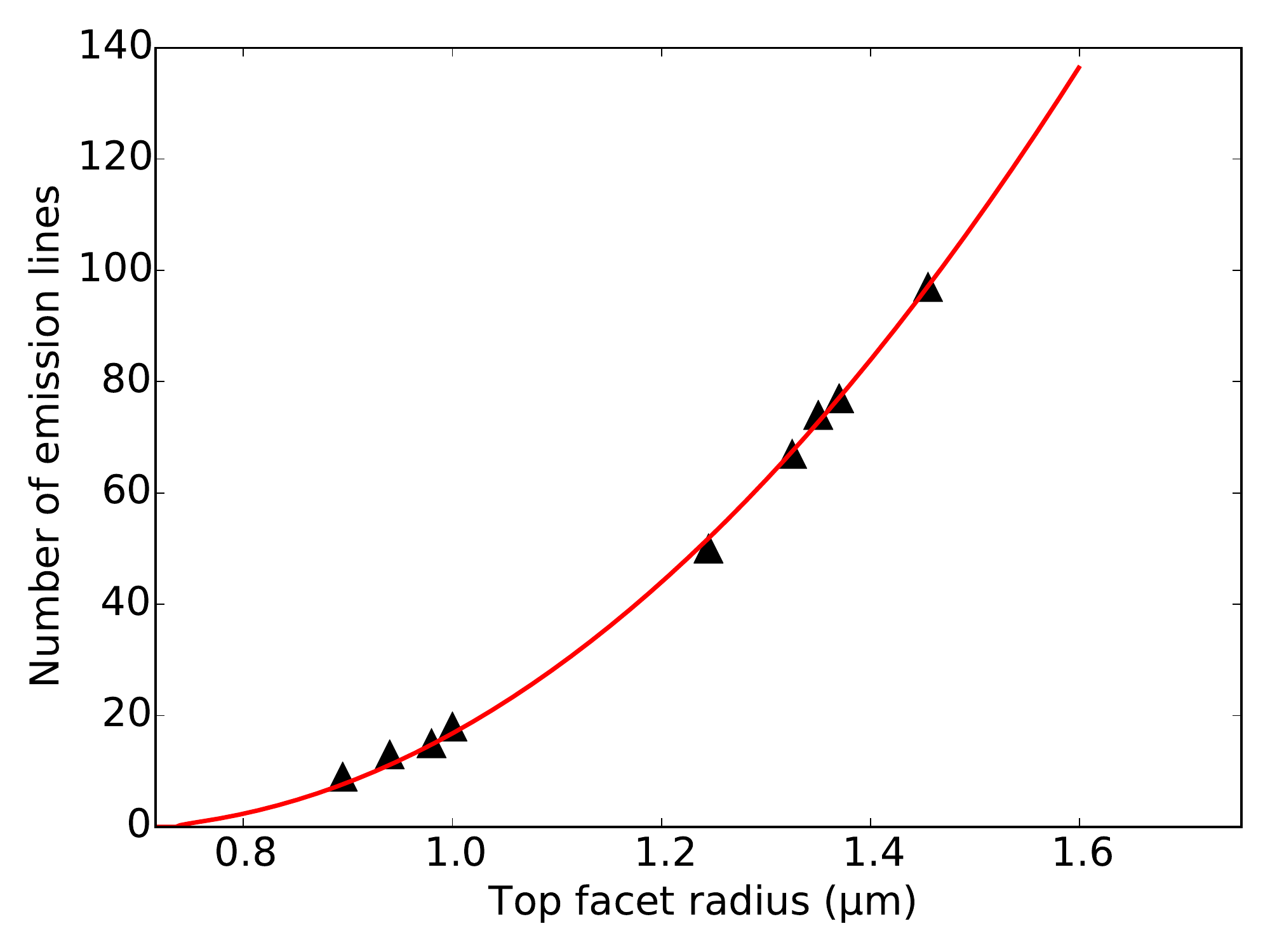}}
 \caption{Number of emission lines within a given spectral range for wires of different diameters. The red line indicates the fit according to Eq. \ref{FitFunction}. The value of $l$ obtained by the fit is \unit{8\pm10}{\nano\metre}.} 
 \label{Stat}
\end{figure}

We considered all emission lines in each PL spectrum and fitted the number of lines as a function of the top facet radius, $R$. It is possible to calculate the shape the cross section at the level of the QDs for any $R$, by using the transformations $X^\prime_0= X_0+(R-R_0)$,  $Y^\prime_0 = Y_0+(R-R_0)$, and $\epsilon^\prime = Y^\prime_0/X^\prime_0$. $R$ is the top radius of a given wire and $R_0$ is the top radius of the one used in our experiment, for which $X_0$ and $Y_0$ are known.

The fit was done using
\begin{align}
N(R) = \varrho\int_0^{2\pi} \frac{\rho(\theta,R,l)^2}{2} \mathrm{d}\theta,
   \label{FitFunction}
\end{align} with $\rho(\theta,R,l)$ being the function that determines the inner boundary at the level of the QDs for a wire of top radius $R$, given a ``dead layer'' of thickness $l$ and a QD surface density $\varrho$.

The two fitting parameters are the emission line surface density $\varrho$ and the maximal radius of a photonic wire that would present no emitting QD, $l$. The fit gives $\varrho=\unit{74\pm2}{lines/\micro m^2}$ and $l=\unit{8\pm10}{\nano\meter}$, with uncertainties corresponding to the standard deviation of each fitted parameter.




\subsection{Probabilistic estimation of position for the QD closest to the edge}
\label{scaling}

\begin{figure}
\includegraphics[width=0.5\textwidth]{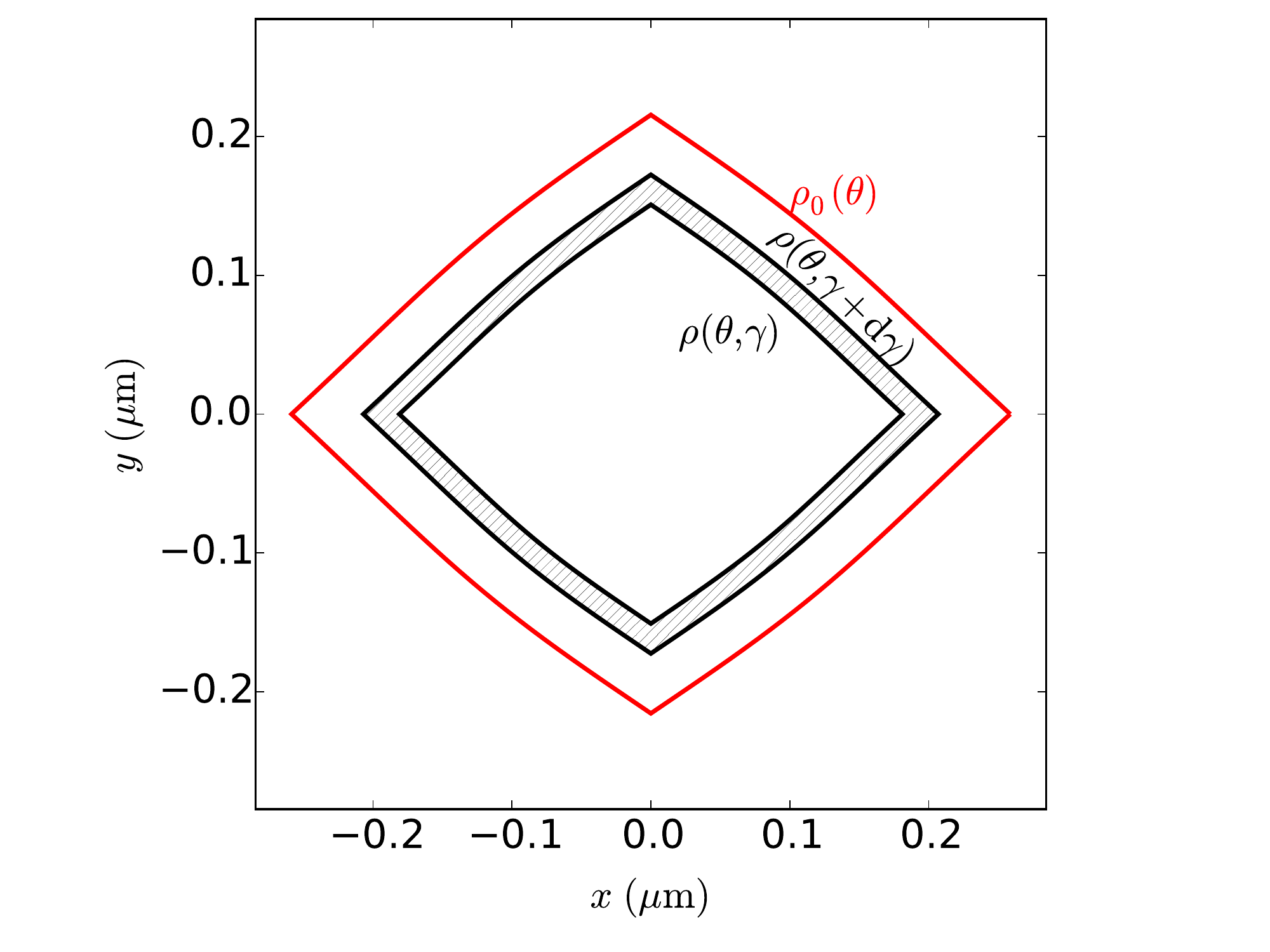}
 \caption{Illustration of the geometry involved in calculating the expected value for the proportionality coefficient $\gamma$ of the bulged lozenge containing the QD furthest from the center of our photonic wire. The boundary of the ``dead layer'' is indicated in red, and described by $\rho_{0}(\theta)$. The region where $N-1$ of the $N$ QDs are located is bounded by  the curve $\rho(\theta,\gamma)=\gamma\rho_0(\theta)$. The $N$-th QD is in the hatched region, between $\rho(\theta,\gamma)$ and $\rho(\theta,\gamma+\mathrm{d}\gamma)$.}
 \label{ExpRmax_Fig}
\end{figure}

In order to determine a scaling factor that properly converts the energy shifts caused by mechanical stress into an $(x,y)$ position, we have estimated,  for $N$ QDs randomly distributed on the optically active area of the wire, the expected position of the QD located closer to the border of the dead layer.
 As Supplementary Fig.\ref{ExpRmax_Fig} shows, this QD, called QD$_m$, will be on a concentric bulged lozenge described by $\rho^\prime(\theta,\gamma)=\gamma\rho_0(\theta)$, where $\rho_0(\theta)$ describes the dead layer border.
 The $N-1$ other QDs are then located inside this lozenge of area $\gamma^2 A_0$, where $A_0=\int_0^{2\pi}(\rho^2_0(\theta)/2) \mathrm{d}\theta$ is the optically active surface of the QD plane (at an elevation$z=0.8\mu$m).
 The probability for a QD to be inside that lozenge is given by the ratio $\gamma^2$ of the areas. The probability
 for  $N-1$ QDs being inside that bulged lozenge is therefore   $\gamma^{2(N-1)}$.

The probability   for finding a QD in the area between bulged lozenges parameterized by $\gamma$ and $\gamma+\mathrm{d}\gamma$, in the hashed region of Supplementary Fig.\ref{ExpRmax_Fig}, is $d\mu(\gamma)=2\gamma\mathrm{d}\gamma$. For finding any of the $N$ QDs in this area the probability is
$d\mu_N(\gamma)=2N\gamma\mathrm{d}\gamma$.
The overall probability  $\mathrm{d}\mathcal{P}(\gamma)$ of finding a QD in this area and all the other QDs inside the region of $\gamma^\prime\leq\gamma$ is then given by
\begin{align}
\mathrm{d}\mathcal{P}(\gamma) = 2N\gamma^{2N-1}\mathrm{d}\gamma .
\label{Pdensity}
\end{align}

We can compute the expected value $\gamma_m$ for the extremal QD, together with  its standard deviation, using the definitions of the expected value and variance:
\begin{eqnarray}
\langle \gamma_m\rangle&=\int_0^1 \gamma\left(2N\gamma^{2N-1}\right)\mathrm{d}\gamma,\nonumber\\
\langle \gamma_m\rangle&=\frac{2N}{2N+1}
\label{average}
\end{eqnarray}
and the corresponding statistical uncertainty
\begin{align}
\delta\gamma_m=\sqrt{\langle \gamma_m^2\rangle -\langle \gamma_m\rangle^2}=\sqrt{\frac{N}{N+1}-\frac{4N^2}{(2N+1)^2}}.
\label{sigma}
\end{align}

Experimentally, we measure the spectral shifts $\Delta\omega_{i,X}$ and $\Delta\omega_{i,Y}$ along the $X$ and $Y$ polarisations for each QD. The QD closest to an edge of the bulged lozenge  will be QD$_m$ and positioned on the $\gamma_m$ lozenge. This will determine the absolute position of QD$_m$  within the wire section. In our case, we find that QD$_m$ is the QD labelled H on the map given in Fig.3 of the main text. It corresponds to   $\rho_m  = 187\pm7$nm.

For the final value of $\rho_m$, we have to account for the dead layer estimation (cf section \ref{Dead_layer} above), and we finally obtain $\rho_m  = 179\pm12$nm.

\subsection{Impact of external stress on the QD emission energy}
\label{impact_stress}

We derive here a simple analytical estimation of the impact of external stress on the emission energy of a self-assembled QD grown along the $z=[001]$ direction. To a very good approximation, the stress response is dominated by single-particle effects, {\it i.e.}, by a modification of single electron and hole energy levels. The QD emission energy can then be expressed as $E_{g,hh} + E_{c,e} + E_{c,hh}$, where $E_{g,hh}$ is the heavy hole (hh) bandgap of the QD material; $E_{c,e}>0$ and $E_{c,hh}>0$ are the energies of the confined electronic and hh levels, measured with respect to the edge of the QD material bands.

We first determine the shift $\Delta E_{g,hh}$ of the {\it hh} bandgap induced by a `small' external stress. In a self-assembled QD, the QD material experiences a large intrinsic bi-axial compressive strain, which strongly lifts the degeneracy of heavy and light hole bands. Following Ref.~\cite{StepanovSI}, we apply first-order perturbation theory to the Bir-Pikus Hamiltonian and obtain $\Delta E_{g,hh} = a \epsilon_h + \frac{1}{2} b\epsilon_{sh}$, with $a = a_c - a_v$ (with the convention $a_c <0$ and  $a_v>0$) and $b$ the deformation potentials of the QD material \cite{Yu,Vurgaftman}; $\epsilon_{h} = \epsilon_{xx}+\epsilon_{yy}+\epsilon_{zz}$ and $\epsilon_{sh} = 2\epsilon_{zz}-\epsilon_{xx}-\epsilon_{yy}$ are respectively the hydrostatic and the tetragonal shear strain induced by the external stress. We use here the cubic basis ($x=[100],y=[010],z=[001]$).

To estimate the shifts $\Delta E_{c,e}$ and $\Delta E_{c,hh}$ of the electron and hole confined levels, we model our flat-lens QD as a quantum well of thickness $t$. For a weakly bounded electronic level, $\Delta E_{c,e} = \Delta \text{CBO} - 2 (\text{CBO} - E_{c,e}) \Big(  \frac{\Delta \text{CBO}}{\text{CBO}} + \frac{\Delta t}{t} \Big)$. A similar expression is obtained for the hole level, by replacing the conduction band offset (CBO) by the valence band offset (VBO), and $E_{c,e}$ by $E_{c,hh}$. CBO, VBO, can be obtained from a modelling of the QD. $\Delta \text{CBO}$ and $\Delta \text{VBO}$ are determined with the deformation potentials: $\Delta \text{CBO} = \Delta a_c \epsilon_h$ and $\Delta \text{VBO} = -\Delta a_v \epsilon_h + \frac{1}{2}\Delta b \epsilon_{sh}$. $\Delta a_c$, $\Delta a_v$ and $\Delta b$ are the difference between the deformation potentials of the barrier and QD materials. Finally, $\frac{\Delta t}{t} = \epsilon_{zz}$.

To obtain numerical values, we consider an In$_{0.6}$Ga$_{0.4}$As/GaAs QD and use the QD energy levels presented in Ref.~\cite{JonsSI}. GaAs is treated as a mechanically isotropic material, with a Poisson ratio $\nu=0.31$ and a Young modulus $Y = 85.9\:  $GPa. A uniaxial stress applied along $z$ yields $\epsilon_{zz} = Y^{-1} \sigma_{zz}$, $\epsilon_h = (1-2\nu) Y^{-1} \sigma_{zz}$ and $\epsilon_{sh} = 2(1+\nu) Y^{-1} \sigma_{zz}$. A uniaxial stress applied along $x$ yields $\epsilon_{zz} = -\nu Y^{-1} \sigma_{xx}$, $\epsilon_h = (1-2\nu) Y^{-1} \sigma_{xx}$ and $\epsilon_{sh}=-(1+\nu) Y^{-1} \sigma_{xx}$.
This analysis yields several important conclusions: (i) For small stresses associated with the wire vibration, the QD emission shift is only due to the diagonal components of the stress tensor ($\sigma_{xx}$, $\sigma_{yy}$ and $\sigma_{zz}$). (ii) For uniaxial stresses applied along the $x$, $y$ and $z$ directions, the QD emission shift is dominated by the shift of the hh bandgap of the QD material. Shifts associated with confinement effects are typically $10$ times smaller. (iii) The tuning slope for uniaxial stress applied along $z$
 is $67\: \mu \text{eV}/\text{MPa}$. It exceeds the one for a uniaxial stress applied along $x$ or $y$ ($18\: \mu \text{eV}/\text{MPa}$), by a factor of $3.7$.
Along the $z$ direction, the effects of $\epsilon_h$ and $\epsilon_{sh}$ add constructively to induce a large hh bandgap shift. In contrast, the effect of $\epsilon_{sh}$ partially cancels the one of $\epsilon_h$ when the stress is applied along $x$ or $y$.

We have used a commercial finite element modelling software to compute   the stress components $\sigma_{xx}$, $\sigma_{yy}$, $\sigma_{zz}$ and obtain the QD energy shift at any location. We find that the typical QD energy shift
for a $1$nm top facet displacement is of $30\mu eV$ for QD located $200$nm away from the zero-stress line. The linearity of the shift with respect to the distance from the zero-stress line is better than $10^{-4}$.

\subsection{Experimental uncertainty  on the spatial coordinates}
\label{errors}

As discussed  in the main text, the $x_i$ coordinate of  the quantum dot QD$_i$  is obtained from the spectral modulation amplitude $\Delta \omega_{i,X}$, via the following formula :
\begin{equation}
x_i=  \frac{s_{m,X}g_m}{s_{i,X}g_i} \mathcal{A}_{i,X} \rho_m(\theta_m),
\label{xi}
\end{equation}
with
\begin{equation}
\mathcal{A}_{i,X}= \frac{\Delta ^0 \omega_{i,X}}{\sqrt{\Delta ^0 \omega_{m,X}^2 + \Delta ^0 \omega_{m,Y}^2 }}
= \frac{\Delta  \omega_{i,X}}{\sqrt{\Delta  \omega_{m,X}^2 + \left( \frac{\mu_\text{X}d_\text{X}}{\mu_\text{Y}d_\text{Y}} \right)^2 \Delta  \omega_{m,Y}^2 }} ,
\end{equation}
so that the variance $\delta x_i^2$ is given by
\begin{equation}
\delta x_i ^2 =  2\mathcal{A}_{i,X} ^2 \rho_m(\theta_m)   ^2
\left( \delta  g_i^2 /g^2 + \delta s_{i,X} ^2  /s_{m,X} ^2  \right)
 +\rho_m(\theta_m)^2 \delta \mathcal{A}_{i,X}^2 +
 \mathcal{A}_{i,X}  ^2  \delta\rho_m(\theta_m)^2 .
 \label{deltaX}
\end{equation}
We know from \cite{StepanovSI} that   $\sqrt{\delta g_i^2 } / g_0 = 13\% $.
From the FEM calculation that we have performed, the uniformity of the gradient is guaranteed up to $\sqrt{\delta s_{i,X} ^2 } /s_{m,X} = 0.01\% $. In the first term of eq. \ref{deltaX}, the factor of $2$ comes from the fact that $\delta  (g_i-g_m)^2 /g^2 = 2 \delta  g_i^2 /g^2$ and $\delta (s_{i,X}- s_{m,X}) ^2  /s_{m,X} ^2 = 2\delta s_{i,X} ^2  /s_{m,X} ^2 $.
We have seen in  section \ref{scaling} above that
 $\delta\rho_m(\theta_m)= 12$nm.

The variance of $\mathcal{A}_{i,X}$ is given by
\begin{eqnarray}
\delta\mathcal{A}_{i,X}^2&=& \frac{1}{\Delta \omega_{m,X}^2 + \left( \frac{\mu_\text{X}d_\text{X}}{\mu_\text{Y}d_\text{Y}} \right)^2\Delta \omega_{m,Y}^2 } \delta \Delta \omega_{i,X}^2
  +\frac{\Delta \omega_{i,X}^2  \Delta \omega_{m,X}^2}{\left( \Delta \omega_{m,X}^2 + \left( \frac{\mu_\text{X}d_\text{X}}{\mu_\text{Y}d_\text{Y}} \right)^2\Delta \omega_{m,Y}^2 \right) ^3} \delta \Delta \omega_{m,X}^2  \nonumber\\
&&
+  \frac{\left( \frac{\mu_\text{X}d_\text{X}}{\mu_\text{Y}d_\text{Y}}\right)^4 \Delta \omega_{i,X}^2 \Delta  \omega_{m,Y}^2 \delta \Delta  \omega_{m,Y}^2
+ \Delta \omega_{i,X}^2 \Delta  \omega_{m,Y}^4 \left( \frac{\mu_\text{X}}{\mu_\text{Y}}\right)^4 \left( \frac{d_\text{X}}{d_\text{Y}}\right)^2  \delta \left( \frac{d_\text{X}}{d_\text{Y}}\right) ^2}
{\left( \Delta  \omega_{m,X}^2 + \left( \frac{\mu_\text{X}d_\text{X}}{\mu_\text{Y}d_\text{Y}}\right)^2 \Delta  \omega_{m,Y}^2 \right) ^3}.
\end{eqnarray}
where $\delta \Delta \omega_{i,X}^2$ is the uncertainty extracted from the fit of the stroboscopic data of Fig.2 of the main text for QD$_i$.  These fits have been realized by taking into account a linear drift of the wire mechanical frequency during the acquisition of the stroboscopic spectra, leading to a linear phase drift.
The quantity $\delta \left( d_\text{X}/d_\text{Y}\right) ^2=0.03^2$ is the uncertainty in the $x$ and $y$ motion amplitude ratio.

\begin{figure}
\resizebox{0.5\textwidth}{!}{\includegraphics{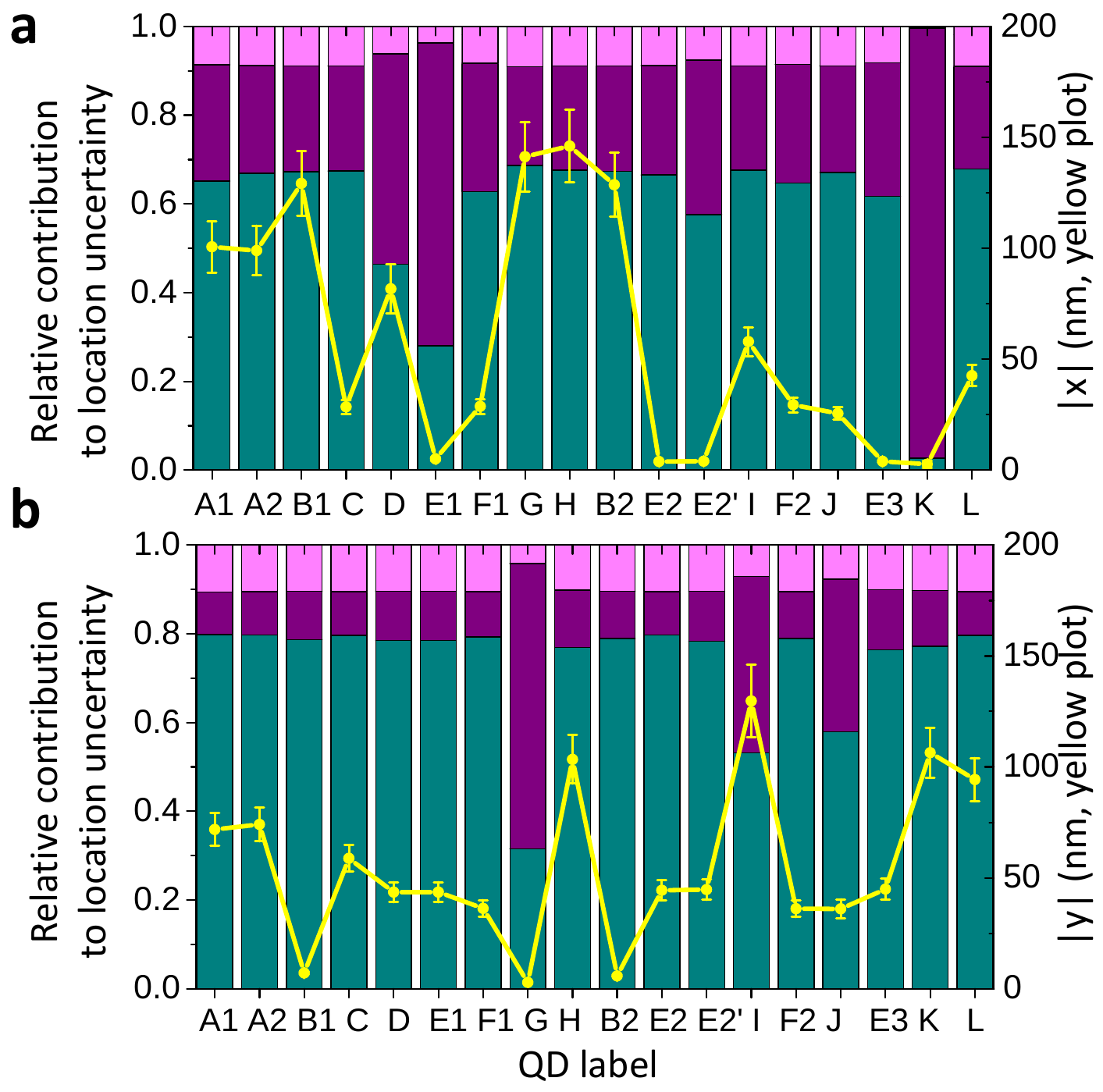}}
 \caption{\textbf{Uncertainty budget}. \textbf{a} is for the $x$ axis, and \textbf{b} is for the $y$ axis. On the left axis, the different  contributions for the total  variance for each line are given according to the following color code : pink, for the uncertainty on the overall scaling factor; purple, for the experimental uncertainty on the line oscillation amplitude; green, for the $g_i$ dispersion; the deviation from a constant gradient is too small to be visible. The right axis on \textbf{a} (resp \textbf{b}) corresponds to the absolute value of the $x$ (resp $y$) coordinate given  by the yellow trace together with the error bars.
 }
 \label{budget}
\end{figure}

The uncertainty budget is given in Supplementary Fig.\ref{budget}. It appears that the main contribution comes from the dispersion on QD responses $g_i$.

\pagebreak


\begin{thebibliography}{0}
\expandafter\ifx\csname natexlab\endcsname\relax\def\natexlab#1{#1}\fi
\expandafter\ifx\csname bibnamefont\endcsname\relax
  \def\bibnamefont#1{#1}\fi
\expandafter\ifx\csname bibfnamefont\endcsname\relax
  \def\bibfnamefont#1{#1}\fi
\expandafter\ifx\csname citenamefont\endcsname\relax
  \def\citenamefont#1{#1}\fi
\expandafter\ifx\csname url\endcsname\relax
  \def\url#1{\texttt{#1}}\fi
\expandafter\ifx\csname urlprefix\endcsname\relax\def\urlprefix{URL }\fi
\providecommand{\bibinfo}[2]{#2}
\providecommand{\eprint}[2][]{\url{#2}}

\end{thebibliography}


\begin{thebibliography}{unsrt}


\bibitem{Rittweger}
Rittweger, E., Han, K.Y., Irvine, S.E.,  Eggeling, C. and  Hell, S. W., STED microscopy reveals crystal colour centres with nanometric resolution.
\textit{Nature Photon.} \textbf{3}, 144-147 (2009)

\bibitem{Bakr}
Bakr, W. S. et al., A quantum gas microscope for detecting single atoms in a Hubbard-regime optical lattice.
\textit{Nature} \textbf{462}, 74 (2009)

\bibitem{Trotta}
Trotta, R. and Rastelli, A., Engineering of quantum dot photon sources via electro-elastic fields.
in \textit{Engineering the Atom-Photon Interaction} (Spinger, 2015)

\bibitem{Yeo}
Yeo, I., et al., Strain-mediated coupling in a quantum dot-mechanical oscillator hybrid system.
\textit{Nature Nanotech.} \textbf{9}, 106 (2014)

\bibitem{Montinaro}
Montinaro, M., W\"{u}st, G., Munsch, M., Fontana, Y., Russo-Averchi, E., Heiss, M., Fontcuberta i Morral, A., Warburton, R. J. and Poggio, M., Quantum Dot Opto-Mechanics in a Fully Self-Assembled Nanowire.
\textit{Nano Lett.} \textbf{14}, 4454-4460 (2014)

\bibitem{Gerard} G\'{e}rard, J.-M. Sermage, B. ,  Gayral, B.,  Legrand, B.,  Costard, E.,  Thierry-Mieg, V.,
\emph{Phys. Rev. Lett} \textbf{81}, 1110 (1998)



\bibitem{Lodahl}
Lodahl, P., Mahmoodian, S. and Stobbe, S., Interfacing single photons and single quantum dots with photonic nanostructures.
\textit{Rev. Mod. Phys.} \textbf{87}, 347 (2015)




\bibitem{Claudon}
Claudon, J., Bleuse, J., Malik, N.S., Bazin, M., Jaffrennou, P., Gregersen, N., Sauvan, C., Lalanne, P., G\'{e}rard, J.M., A highly efficient single-photon source based on a quantum dot in a photonic nanowire.
\textit{Nature Photon.} \textbf{4}, 174-177 (2010)

\bibitem{Munsch}
Munsch, M., et al., Dielectric GaAs Antenna Ensuring an Efficient Broadband Coupling between an InAs Quantum Dot and a Gaussian Optical Beam.
\textit{Phys. Rev. Lett.} \textbf{110} 177402 (2013)

\bibitem{Dousse}
Dousse, A. et al., Ultrabright source of entangled photon pairs.
\textit{Nature} \textbf{466}, 217 (2010)

\bibitem{Ding}
Ding, X. et al., On-Demand Single Photons with High Extraction Efficiency and Near-Unity Indistinguishability from a Resonantly Driven Quantum Dot in a Micropillar. \textit{Phys. Rev. Lett.} \textbf{116}, 020401 (2016)

\bibitem{Besombes}
Besombes, L., et al., Probing the Spin State of a Single Magnetic Ion in an Individual Quantum Dot.
\textit{Phys. Rev. Lett.} \textbf{93}, 207403 (2004)


\bibitem{Awschalom}
Awschalom, D. D. et al., Quantum Spintronics: Engineering and Manipulating Atom-Like Spins in Semiconductors.
\textit{Science} \textbf{339}, 1174 (2013)

\bibitem{DeGreve}
De Greve, K. et al. Quantum-dot spin-photon entanglement via frequency down conversion to telecom wavelength.
\textit{Nature} \textbf{491}, 421 (2012)

\bibitem{Gao}
Gao, W. B. et al. Observation of entanglement between a quantum dot spin and a single photon.
\textit{Nature} \textbf{491}, 426 (2012)

\bibitem{Wieck_2005}
Unold, T. et al., Optical Control of Excitons in a Pair of Quantum Dots Coupled by the Dipole-Dipole Interaction.
\textit{Phys. Rev. Lett.} \textbf{94}, 137404 (2005)

\bibitem{Savona_2005}
Parascandolo, G. and Savona, V., Long-range radiative interaction between semiconductor quantum dots.
\textit{Phys. Rev. B} \textbf{71}, 045335 (2005)


\bibitem{Kasprzak}
Kasprzak, J., Patton B., Savona V., and Langbein W., Coherent coupling between distant excitons revealed by two-dimensional nonlinear hyperspectral imaging.
\textit{Nature Photon.} \textbf{5}, 123 (2011)

\bibitem{Mermillod}
Mermillod, Q., Jakubczyk, T., Delmonte, V., Delga, A., Peinke, E., G\'{e}rard, J.-M.,  Claudon, J. and  Kasprzak, J., Harvesting, Coupling, and Control of Single-Exciton Coherences in Photonic Waveguide Antennas.
\textit{Phys. Rev. Lett.} \textbf{116}, 163903 (2016)

\bibitem{Temnov}
Temnov, V. V. and Woggon U., Superradiance and Subradiance in an Inhomogeneously Broadened Ensemble of Two-Level Systems Coupled to a Low-Q Cavity.
\textit{Phys. Rev. Lett.} \textbf{95}, 243602 (2005)

\bibitem{Auffeves}
Auff\`{e}ves, A., Gerace, D., Portolan, S., Drezet, A., and Fran\c{c}a Santos, M., Few emitters in a cavity: from cooperative emission to individualization.
\textit{New J. Phys.} \textbf{13}, 093020 (2011)

\bibitem{Betzig_NSOM2}
Betzig, E.,  and Chichester, R. J., Single Molecules Observed by Near-Field Scanning Optical Microscopy.
\textit{Science} \textbf{262}, 1422-1425 (1993)


\bibitem{Matsuda}
Matsuda, K. et al., Near-field optical mapping of exciton wave functions in a GaAs quantum dot.
\textit{Phys. Rev. Lett. \textbf{91}}, 177401 (2003)

\bibitem{Brown} Brown, R. W., Norman Cheng, Y.-C., Haacke, E. M., Thompson, M. R.,  and Venkatesan, R., \textit{Magnetic Resonance Imaging: Physical Principles and Sequence Design} (John Wiley \& Sons, New York City, 2014)

\bibitem{Mamin}
Mamin, H.J. et al., Nanoscale Nuclear Magnetic Resonance with a Nitrogen-Vacancy Spin Sensor.
\textit{Science} \textbf{339}, 557 (2013)

\bibitem{Degen}
Degen, C.L., Poggio, M., Mamin, H.J., Rettner, C.T., and Rugar, D., Nanoscale magnetic resonance imaging.
\textit{Proc. Nat. Acad. Sci.} \textbf{106}, 1313 (2009)

\bibitem{Sanii}
Sanii, B., and Ashby, P.D., High Sensitivity Deflection Detection of Nanowires.
\textit{Phys. Rev. Lett.} \textbf{104}, 147203 (2010)

\bibitem{Gloppe}  Gloppe, A.,	Verlot, P.,	Dupont-Ferrier,	E., Siria, A.,	Poncharal, P.,	Bachelier, G.,	 Vincent, P., and  Arcizet, O.,
Bidimensional nano-optomechanics and topological backaction in a non-conservative radiation force field,
        Nature Nanotechnology \textbf{9}, 920, (2014)


\bibitem{Stepanov}
Stepanov, P.,   Elzo-Aizarna, M.,  Bleuse, J.,  Malik, N.S.,  Cur\'{e}, Y.,  Gautier, E.,
 Favre-Nicolin, V.,  G\'{e}rard, J.M., and  Claudon, J., Large and Uniform Optical Emission Shifts in Quantum Dots Strained along Their Growth Axis.
\textit{Nano Lett.} \textbf{16}, 3215-3220 (2016)

\bibitem{Zhiming}
Wang, Z. M., \textit{Self-Assembled Quantum Dots} (Springer-Verlag, New York City, 2008)

\bibitem{Kuklewicz} Kuklewicz, C. E., Malein, R. N. E., Petroff, P. M., and Gerardot B. D., Electro-Elastic Tuning of Single Particles in Individual Self-Assembled Quantum Dots. \textit{Nano Lett.} \textbf{12}, 3761 (2012)

\bibitem{Wu} Wu, X., Dou, X., Ding K., Zhou, P., Ni, H., Niu, Z., Jiang, D., and Sun B., In situ tuning the single photon emission from single quantum dots through hydrostatic pressure. \textit{Appl. Phys. Lett.} \textbf{103}, 252108 (2013)


\bibitem{Teissier} Teissier, J., Barfuss, A., Appel, P., Neu, E., Maletinsky, P., Strain Coupling of a Nitrogen-Vacancy Center Spin to a Diamond Mechanical Oscillator, \emph{Phys. Rev. Lett.} \textbf{113}, 020503 (2014)

\bibitem{Thorpe}  Thorpe, M.J.,  Rippe, L.,  Fortier, T.M., Kirchner, M.S., Rosenband, T.,
Frequency stabilization to $6\times10^{-16}$ via spectral-hole burning, \textit{Nature Photon.}, \textbf{5}, 688 (2011)


\bibitem{Yeo_2011}
Yeo, I., et al., Surface effects in a semiconductor photonic nanowire and spectral stability of an embedded single quantum dot.
\textit{Appl. Phys. Lett.} \textbf{99}, 233106 (2011)

\end{thebibliography}

\begin{thebibliography}{unsrt}




\bibitem{StepanovSI}
Stepanov, P.,   Elzo-Aizarna, M.,  Bleuse, J.,  Malik, N.S.,  Cur\'{e},  Y.,  Gautier, E.,
 Favre-Nicolin, V.,  G\'{e}rard, J.M., and  Claudon, J., Large and Uniform Optical Emission Shifts in Quantum Dots Strained along Their Growth Axis.
\textit{Nano Lett.} \textbf{16}, 3215 (2016)

\bibitem{Yu}Yu, P.Y., and Cardona, M., Fundamentals of semiconductors. Springer ed. (2010)

\bibitem{Vurgaftman}
 Vurgaftman, I.,  Meyer, J.R.,  Ram-Mohan, L.R.,
Band parameters for III-V compound semiconductors and their alloys,
\emph{J. Appl. Phys. } \textbf{89}, 5815 (2001)

\bibitem{JonsSI} J\"{o}ns, K.D., Hafenbrak, R.,  Singh, R.,  Ding, F., Plumhof, J.D., Rastelli, A., Schmidt, O.G.,  Bester, G., and  Michler, P., Dependence of the Redshifted and Blueshifted Photoluminescence Spectra of Single In$_x$Ga$_{1-x}$As/GaAs Quantum Dots on the Applied Uniaxial Stress. \textit{Phys. Rev. Lett.} \textbf{107}, 217402 (2011)

\end{thebibliography}
\end{document}